\documentclass[aps,prx,reprint,longbibliography,floatfix]{revtex4-2}
\usepackage{amsmath,amssymb,graphicx}
\usepackage{bm}
\usepackage{soul}
\usepackage[colorlinks, linkcolor= blue, citecolor = blue, urlcolor=blue]{hyperref}
\usepackage{makecell}
\usepackage{physics}
\usepackage{appendix}
\usepackage{ragged2e}
\usepackage{multirow}
\usepackage{array}
\usepackage{braket}
\usepackage[useregional]{datetime2}
\def\be{\begin{equation}}
\def\ee{\end{equation}}
\def\bea{\begin{eqnarray}}
\def\eea{\end{eqnarray}}
\def\nn{\nonumber}
\begin{document}
\title{Electric-Polarization Probe of the Magnon Orbital Moment Current in Altermagnet}
\author{Sankar Sarkar}
\email{sankars24@iitk.ac.in}
\affiliation{Department of Physics, Indian Institute of Technology Kanpur, Kanpur-208016, India}
\author{Amit Agarwal}
\email{amitag@iitk.ac.in}
\affiliation{Department of Physics, Indian Institute of Technology Kanpur, Kanpur-208016, India}
%
%
\begin{abstract}
Efficient transport of spin and orbital moments, and their electrical detection, are among the main challenges in spintronics and orbitronics. In magnetic insulators, these currents are mediated by magnons. In addition to carrying spin and orbital moment, the orbital motion of a magnon combined with its magnetic moment,  generates an effective electric dipole moment. Here, we develop a theoretical framework for Seebeck- and Nernst-type transport of the magnon orbital moment (MOM) and its associated electric dipole moment (EDM). We identify a Drude-like scattering contribution and an intrinsic component governed by the generalized Berry curvatures of magnon bands. We show that a measurable transverse voltage generated by the EDM current provides a direct electrical detection scheme for magnon orbital transport. Applying our theory to an hexagonal altermagnet, we obtain an experimentally accessible voltage of approximately $0.4~\mu$V. Our results establish a concrete electrical probe of magnon orbital transport and highlight magnons as potential low-dissipation information carriers for orbitronics. 
%
%
%
\end{abstract}
\maketitle
%
%
%
\section{Introduction}
The need to store and process information faster than conventional charge-based electronics has driven the rapid development of spintronics~\cite{Wolf_2001, Sarma_2004} and spin-caloritronics~\cite{bauer_2012, boona_2014}. Spintronics exploits spin-dependent phenomena such as the spin Hall effect~\cite{Dyakonov_1971, Hirsch_1999, kato_2004, Sinova_2004, Wunderlich_2005, Sinova_2015}, Edelstein effect~\cite{EDELSTEIN_1990, Kato_2004_, Gorini_2017, Sarkar_2025}, and spin current-driven spin-orbit torque \cite{Takahashi_2008, Manchon_2019, Shao_2021}. In contrast, spin-caloritronics focuses on thermally driven responses, including the spin Seebeck effect~\cite{uchida_2008, jaworski_2010, Minggang_2011, sankar_2025_2}, spin Nernst effect~\cite{meyer_2017, sheng_2017, sankar_2025_2}, and thermal spin magnetization~\cite{WANG_2010, Shitade_2019, sankar_2025_2}. Parallel to these advances, the transport of angular momentum carried by orbital degrees of freedom has given rise to the emerging field of orbitronics. Recent studies have revealed sizable orbital angular momentum in Bloch electrons~\cite{Tanaka_2008, Kontani_2009, Dongwook_2018, Kamal_2021,  burgos_2024, Koushik_2026}, enabling orbital analogs of spin phenomena such as the orbital Edelstein effect~\cite{Yoda_2018, Pratap_2024}, orbital Hall effect~\cite{Tanaka_2008, Kontani_2009, Dongwook_2018,  choi_2023}, and orbital Nernst effect~\cite{Salemi_2022}.

Magnons, the elementary excitations of magnetically ordered states, provide an alternative route for transporting orbital moments. Unlike electrons, magnons are charge-neutral bosonic quasiparticles that propagate without Joule heating~\cite{Baltz_2018}. This makes them attractive candidates for energy-efficient information processing and quantum technologies \cite{chumak_2015}. However, their charge neutrality poses a fundamental challenge: how to experimentally detect the transport of MOM? Addressing this question requires both a consistent transport theory and a viable electrical detection scheme. Recent works have begun exploring MOM transport~\cite{Gyungchoon_2024, Daehyeon_2025, Yannick_2025, Quang_2024} and its connection to magnon orbital-motion-induced electric polarization (density of EDMs), which offers a potential route for electrical detection~\cite{Gyungchoon_2024, Neumann_2024, Quang_2024}. In particular, Ref.~\cite{Quang_2024} developed a theory of the MOM Nernst effect and quantified the voltage associated with equilibrium polarization. However, the role of non-equilibrium electric polarization generated by thermally driven EDM currents has not yet been addressed. 

\begin{figure}
    \centering
    \includegraphics[width=\linewidth]{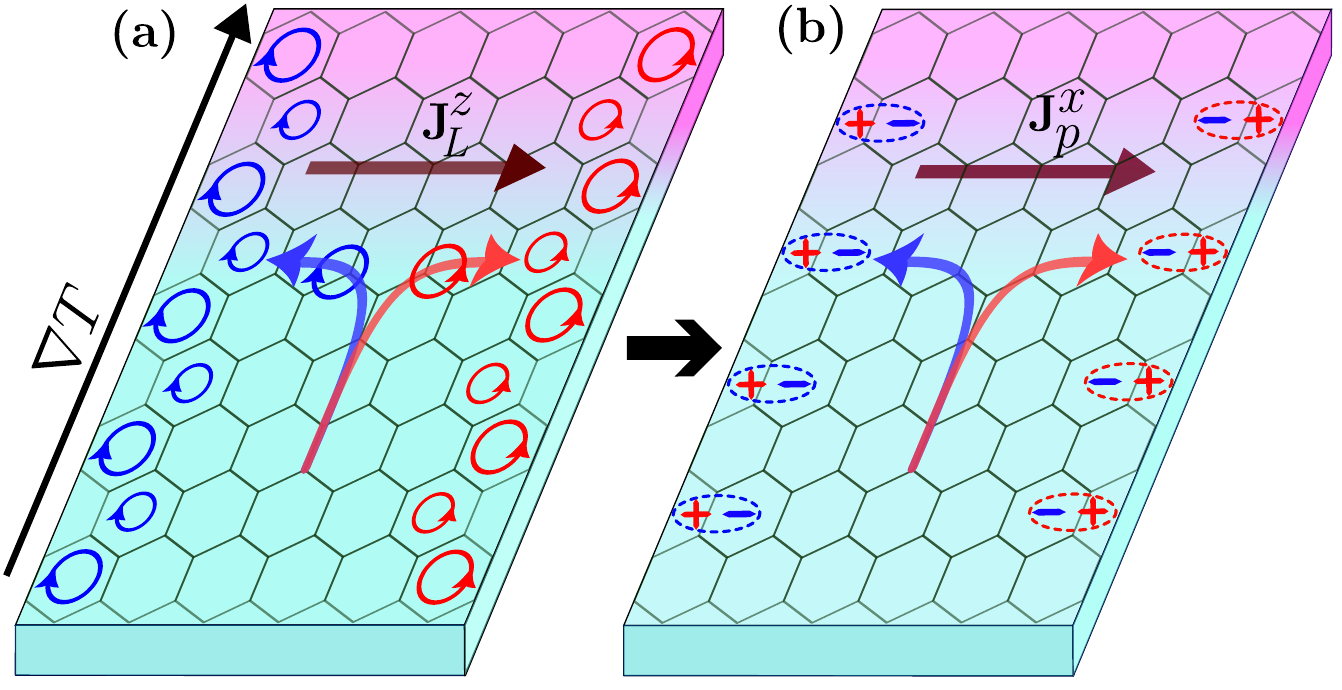}
    \caption{\justifying {\bf MOM and EDM Nernst responses in an hexagonal altermagnet}.  
    (a) MOM Nernst effect: oppositely oriented orbital moments flow in opposite transverse directions under a thermal gradient and accumulate at the sample edges. 
    (b) The associated EDM Nernst effect: the transverse flow of EDMs leads to electric polarization and edge accumulation. This enables electrical detection of magnon orbital transport.
    \label{fig:Fig_1} }
\end{figure}
In this work, we develop a density-matrix-based quantum kinetic framework to address the experimental detection of magnon orbital transport. We define the MOM operator as 
\( \hat{\bm L} = (\hat{\bm r} \times \hat{\bm v} - \hat{\bm v} \times \hat{\bm r})/4 \), 
where $\hat{\bm r}$ and $\hat{\bm v}$ denote the position and velocity operators, respectively~\cite{Gyungchoon_2024}. Despite their charge neutrality, the orbital motion of magnons together with their intrinsic magnetic moment $\hat{\bm m}$, generates an effective EDM~\cite{Neumann_2024, Quang_2024},
\( \hat{\bm p} = (\hat{\bm v} \times \hat{\bm m} - \hat{\bm m} \times \hat{\bm v})/2c^2 \), with $c$ being the speed of light. The density of these EDMs produces an electric polarization, offering a direct route for electrical detection. Treating MOM and EDM on equal footing, we derive unified expressions for their Seebeck- and Nernst-type responses [see Fig.~(\ref{fig:Fig_1})]. We decompose these responses into Fermi-surface (Drude-like) and Fermi-sea (intrinsic) contributions. In the weak-scattering limit, we show that the intrinsic Fermi-sea response is governed by generalized Berry curvatures, namely the orbital Berry curvature (OBC) and dipolar Berry curvature (DBC). We further demonstrate that the interplay between non-equilibrium EDM currents and equilibrium polarization generates a measurable transverse voltage. 
To illustrate our framework, we apply it to a hexagonal  altermagnet~\cite{Smejkql_2022, camerano_2025, Nirmalya_2025, Sankar_2025}. For realistic thermal gradients, the predicted voltage reaches approximately $0.4~\mu$V, which lies within current experimental resolution. These results establish a concrete electrical detection protocol for magnon orbital transport.

\section{Theory of MOM and EDM transport}
In this section, we develop the theoretical framework for temperature-gradient driven transport of MOM and its associated EDM within a density-matrix-based quantum kinetic approach. Magnons, as bosonic excitations of magnetically ordered systems, obey the Bose–Einstein distribution, 
$f(\epsilon) = [ \text{exp}( \epsilon/k_B T ) - 1]^{-1}$, 
where $\epsilon$ is the magnon energy, $T$ is the temperature, and $k_B$ is the Boltzmann constant. Since the magnon number is not conserved, the chemical potential is set to zero.

Unlike electronic systems, the formulation of magnon transport requires special care in choosing the appropriate basis. The magnon Hamiltonian $\mathcal{H}_{\bm{k}}$ cannot be diagonalized by a simple unitary transformation, as this would not preserve the bosonic commutation relations. Instead, a para-unitary Bogoliubov transformation has to be employed (see Sec.~\ref{sea:model_calculation} for details). 
To properly account for the bosonic particle–hole structure, we work with the pseudo-Hermitian Bloch Hamiltonian, $ 
\bar{\mathcal{H}}_{\bm{k}} = \sigma^3 \mathcal{H}_{\bm{k}}, 
$
which satisfies the pseudo-Hermiticity condition $
\sigma^3 \bar{\mathcal{H}}_{\bm{k}}^{\dagger} \sigma^3 = \bar{\mathcal{H}}_{\bm{k}}
$. Here, $\sigma^3 = \mathrm{diag}(1,1,-1,-1)$ acts in particle–hole space. The right and left eigenstates of $\bar{\mathcal{H}}_{\bm{k}}$ are defined by
\[
\bar{\mathcal{H}}_{\bm{k}} \ket{u^R_{n\bm{k}}} = \bar{\varepsilon}_{n\bm{k}} \ket{u^R_{n\bm{k}}}, 
\quad 
\bra{u^L_{n\bm{k}}} \bar{\mathcal{H}}_{\bm{k}} = \bar{\varepsilon}_{n\bm{k}} \bra{u^L_{n\bm{k}}},
\]
with $\bra{u^L_{n\bm{k}}} = \bra{u^R_{n\bm{k}}} \sigma^3$ \cite{Gyungchoon_2024, Daehyeon_2025, Quang_2024}. For notational simplicity, we denote $\ket{u^R_{n\bm{k}}}$ by $\ket{u_{n\bm{k}}}$ in the following. The eigenstates satisfy the generalized normalization and completeness relations
\(
\bra{u_{n\bm{k}}} \sigma^3 \ket{u_{m\bm{k}}} = \sigma^3_{nn}\delta_{nm}\), and 
\( \mathcal{I} = \sum_m \sigma^3_{mm} \ket{u_{m\bm{k}}} \bra{u_{m\bm{k}}} \sigma^3,
\) which ensure consistency with bosonic commutation relations. 
\subsection{First order density matrix}
Using the equilibrium distribution and Bloch states introduced above, we now analyze how a temperature gradient modifies the magnon distribution. The effect of the thermal field is incorporated through the magnon density matrix $\rho(\bm k, t)$, which evolves according to the quantum Liouville equation \cite{Sekine_2020, Harsh_2023, Sankar_2025}
\begin{equation}
\frac{\partial \rho (\bm k, t)}{\partial t} 
+ \frac{i}{\hbar} [\bar{\mathcal{H}}_{\bm k}, \rho] 
+ \frac{\rho (\bm k, t)}{\tau} 
= \mathcal{D}_T[\rho (\bm k, t)]~.
\label{quantum_lioiville_equation}
\end{equation}
Here, $\tau$ is a phenomenological relaxation time incorporating magnon scattering processes. For simplicity, we assume $\tau$ to be momentum independent, which does not affect the qualitative behavior of the response.

Treating the temperature gradient within Luttinger’s gravitational-potential formalism \cite{Luttinger_1964, Tatara_2015}, the thermal driving term is
\be
\mathcal{D}_T[\rho] = - \frac{1}{2 \hbar} \bm{E}_T \cdot \big[ \{ \bar{\mathcal{H}}_{\bm k}, \partial_{\bm k} \rho \} - i [ \bm{\mathcal{ R}}, \{ \bar{\mathcal{H}}_{\bm k}, \rho \} ] \big] ~, \label{thermal_driven_term}
\ee
where $\bm{E}_T = -\nabla T/T$ is the thermal field, $\partial_{\bm k} \equiv \partial / \partial \bm k$, and $\bm{\mathcal{R}}$ denotes the Berry connection. The magnon density matrix also satisfies the pseudo-Hermiticity condition $\sigma^3 \hat{\rho}^\dagger \sigma^3 = \hat{\rho}$. 
The equilibrium density matrix consistent with this structure is
\(
\hat{\rho}^{(0)} = \sum_n 
f_n(\bar{\varepsilon}_{n\bm k})
\, |u_{n\bm k}\rangle 
\langle u_{n\bm k}| \sigma^3~,
\)
which is diagonal in band space, 
\(
\rho^{(0)}_{mq}
= \langle u_{m\bm k} | \sigma^3 \hat{\rho}^{(0)} | u_{q\bm k} \rangle
= f_m \delta_{mq}~.
\)

For a dc thermal field, we impose the steady-state condition $\partial_t \rho = 0$. In the weak-field limit, we expand the density matrix perturbatively as
\(
\rho(\bm k) = \rho^{(0)} + \rho^{(1)} + \cdots ,
\)
where $\rho^{(1)} \propto |\bm E_T|$ is the first-order correction. Substituting this expansion into Eq.~\eqref{quantum_lioiville_equation}, we obtain (see Appendix~\ref{details_of_first_order_DM} for detailed calculations)
\be\label{first_order_dm}
\rho^{(1)}_{nm} 
= - \frac{E_T^a}{\hbar}
\left[ \delta_{nm} \tau \bar{\varepsilon}_{n \bm{k}} \partial_{k_a} f_n
+ i g_{nm} \mathcal{R}^a_{nm} \xi_{nm}
\right]~.
\ee
Here, we have defined  $g_{nm} = [1/\tau + i \bar{\omega}_{nm}]^{-1}$, $\xi_{nm} = ( \sigma^3_{nn} \bar{\varepsilon}_{n \bm k} f_n - \sigma^3_{mm} \bar{\varepsilon}_{m \bm k} f_m )$ and $\bar{\omega}_{nm} = (\bar{\varepsilon}_{n \bm k} - \bar{\varepsilon}_{m \bm k} )/\hbar$. Repeated Roman indices are summed over.
The interband Berry connection is defined as \cite{Koyama_2025}
\[
\mathcal{R}^a_{nm} 
= \sigma^3_{nn}
\langle u_{n\bm k} | i \sigma^3 | \partial_{k_a} u_{m\bm k} \rangle .
\]
In Eq.~\eqref{first_order_dm}, the first term describes the intraband (Drude-like) response proportional to the relaxation time, while the second term captures interband coherence effects governed by Berry connection.
Since we focus on linear response, we neglect higher-order corrections to the density matrix.

To characterize the band geometry governing intrinsic responses, we introduce the quantum geometric tensor \cite{Harsh_2023}, 
\(
{\mathcal Q}^{ab}_{nm} = \mathcal{R}^a_{nm} \mathcal{R}^b_{mn},
\)
whose real and imaginary parts correspond to the quantum metric and Berry curvature, respectively. For our case, the quantum metric and the Berry curvature are explicitly given by, 
\begin{align}
\mathcal{G}^{ab}_n 
&= \Re \sum_{m \neq n}
\sigma^3_{nn} \sigma^3_{mm}
\frac{v^a_{nm} v^b_{mn}}{\bar{\omega}_{nm}^2}, \\
\Omega^{ab}_n 
&= -2 \Im \sum_{m \neq n}
\sigma^3_{nn} \sigma^3_{mm}
\frac{v^a_{nm} v^b_{mn}}{\bar{\omega}_{nm}^2},
\end{align}
where we used
\(
\sigma^3_{nn} v^a_{nm}
= i \bar{\omega}_{nm} \mathcal{R}^a_{nm}
\)
for $n \neq m$. The interband velocity along the $a$ direction is given by $ v^a_{nm} = (1/\hbar) \bra{u_{n \bm k}} \partial_{k_a} \mathcal{H}_{\bm k} \ket{u_{m \bm k}} $. The effect of Berry curvature on the magnon Nernst effect is well studied in the literature \cite{Matsumoto_2011_prl, Matsumoto_2011, Zyuzin_2016, Sankar_2025}. In the following section, we define the generalized version of these band geometric quantities and explore their effect on the transport properties of MOM and EDM in the presence of a thermal field.

\subsection{MOM and EDM current response tensors}
Using the first-order density matrix obtained above, we now construct the linear response theory for the transport of the MOM and its associated EDM. The formalism developed below applies equally to both observables. Although magnons are charge-neutral excitations of a magnetically ordered state, they carry an intrinsic magnetic moment \cite{Neumann_2020}. For a collinear antiferromagnet with conserved spin, the magnetic moment operator is given by \cite{Zyuzin_2016, Cheng_2016}
$
\hat{\bm m} = g \mu_B \hat{\bm s},
$
where $\hat{\bm s} = \pm \hat{z}$ denotes the spin polarization of the magnon band and $\mu_B$ is the Bohr magneton. 

Relativistic electrodynamics implies that the orbital motion of a magnetic moment generates an effective electric dipole moment. The corresponding EDM operator is
\begin{equation}
\hat{\bm p}
=
\frac{1}{2c^2}
\left(
\hat{\bm v} \times \hat{\bm m}
-
\hat{\bm m} \times \hat{\bm v}
\right),
\end{equation}
where $\hat{\bm v} = (1/\hbar)\partial_{\bm k}\mathcal{H}_{\bm k}$ is the magnon velocity operator in the laboratory frame and $c$ is the speed of light \cite{Katsura_2005, Liu_2011, Neumann_2024}. Ref.~\cite{Quang_2024} analyzed the intrinsic contribution to the EDM Nernst response. However, a comprehensive transport framework incorporating both intrinsic and scattering-induced contributions to EDM transport has not yet been formulatd. The translational motion of these EDMs, driven by a temperature gradient, leads to their accumulation at the sample edges. This provides an electrical detection mechanism of MOM transport and its edge accumulations.  

In the following, we treat the MOM operator $\hat{\bm L}$ and the orbital-motion-induced EDM operator $\hat{\bm p}$ on equal footing. To this end, we introduce a generalized current operator \cite{Gyungchoon_2024, Quang_2024}
\begin{equation}
\hat{\mathcal{J}}^{\gamma ; a}_{\mathcal{O}}
=
\frac{1}{4}
\left(
\hat{\mathcal{O}}^{\gamma} \sigma^3 \hat{v}^a
+
\hat{v}^a \sigma^3 \hat{\mathcal{O}}^{\gamma}
\right),
\end{equation}
where $\hat{\mathcal{O}}^{\gamma} = (\hat{L}^{\gamma}, \hat{p}^{\gamma})$. The superscripts $\gamma$ and $a$ denote the component of the observable and the direction of the current, respectively. Using the first-order density matrix from Eq.~(\ref{first_order_dm}), the linear response current is obtained from the trace of the density matrix with the generalized current operator \cite{Xiao_2006, Harsh_2023, Varshney_2023}. We obtain, 
\begin{equation}
J^{\gamma; a}_{\cal O} = \sum_{\bm k, n, m} \sigma^3_{mm}
\rho^{(1)}_{nm} \mathcal{J}^{\gamma; a}_{\mathcal{O}; mn}
+ \sum_{\bm k, n} \left[ \bm E_T \times \bm M^{\gamma}_{\mathcal{O}; n} \right]^a.
\label{general_current}
\end{equation}
Here, $\sum_{\bm k} \equiv \int d^d k /(2\pi)^d$, with $d$ being the spatial dimension of the system. The interband matrix elements of the currents are defined as
$ \mathcal{J}^{\gamma;a}_{\mathcal{O}; mn} = \langle u_{m\bm k}| \hat{\mathcal{J}}^{\gamma;a}_{\mathcal{O}} |u_{n\bm k}\rangle$.  
The first term in Eq.~(\ref{general_current}) captures the density-matrix-induced transport response, while the second term represents the orbital magnetization correction required for thermodynamically consistent transport theory. In the appropriate limit, this expression reproduces the previously established orbital Nernst coefficient \cite{Gyungchoon_2024, Quang_2025, Daehyeon_2025, Quang_2024}.
The generalized Berry-curvature-induced orbital magnetization 
$ M^{\gamma;c}_{\mathcal{O}; n} $ is given by
\be
M^{\gamma; c}_{ \mathcal{O}; n} (\bm k) = \frac{1}{2\hbar} \epsilon_{cab}~ c_1(f_n)  ~ \Omega^{\gamma;ab}_{\mathcal{O}; n} (\bm k).
\ee
Here, $c_1(f_n) = k_B T \ln(1 - e^{-\beta \bar{\varepsilon}_{n \bm k}}) = - k_B T \ln(1 + f_n) $ and $\Omega^{\gamma; ab}_{\mathcal{O}; n}$ is the generalized Berry curvature for $n$th band defined in Eq.~\eqref{generalized_QGT}. 

In the linear-response regime, the MOM and EDM currents take the form $ J^{\gamma;a}_{\mathcal{O}} = - \frac{\nabla^b T}{T} \, \alpha^{\gamma;ab}_{\mathcal{O}}$, 
where $\alpha^{\gamma;ab}_{\mathcal{O}}$ is the response tensor. Here, $a$ denotes the direction of the current and $b$ the direction of the applied thermal gradient.
The response tensor is given by
\begin{align}
\alpha^{\gamma;ab}_{\mathcal{O}} &= - \sum_{\bm k, n} \Big[\tau\, \mathcal{J}^{\gamma;a}_{\mathcal{O};nn}
\, \bar{\varepsilon}_{n \bm k} \, v^b_{nn} \frac{\partial f_n}{\partial \bar{\varepsilon}_{n \bm k}} - \epsilon_{abc} M^{\gamma;c}_{\mathcal{O}; n} \notag\\
&\qquad + \frac{1}{\hbar} \sum_{m \neq n} g_{nm}\, \bar{\omega}_{nm}\, \mathcal{S}^{\gamma;ab}_{\mathcal{O}; mn}\, \xi_{nm} \Big]. \label{MOMCC}
\end{align}
The first term represents the intraband (Drude-like) contribution proportional to the relaxation time $\tau$, the second term corresponds to the orbital magnetization correction, and the third term captures interband coherence effects governed by band geometry.
The third term naturally introduces a generalized quantum geometric tensor \cite{sankar_2025_2},
\begin{align}
\mathcal{S}^{\gamma;ab}_{\mathcal{O}; mn} = \sigma^3_{mm} \sigma^3_{nn} \frac{\mathcal{J}^{\gamma;a}_{\mathcal{O}; mn}
\, v^b_{nm}}{\bar{\omega}_{mn}^2}= \mathcal{G}^{\gamma;ab}_{\mathcal{O}; mn} -\frac{i}{2} \Omega^{\gamma;ab}_{\mathcal{O}; mn},
\label{generalized_QGT}
\end{align}
whose real and imaginary parts correspond to the generalized quantum metric and generalized Berry curvature, respectively.

To understand the physical origin of the responses, we decompose the total coefficient into Fermi-surface and Fermi-sea contributions. From Eq.~(\ref{MOMCC}), the first term contains the derivative of the Bose–Einstein distribution and is therefore analogous to a Fermi-surface contribution in electronic transport. In contrast, the second and third terms involve the distribution function itself and therefore correspond to Fermi-sea contributions. Accordingly, we write $
\alpha^{\gamma; ab}_{\mathcal{O}} = \alpha^{\gamma;ab}_{\mathcal{O};\rm Surf} + \alpha^{\gamma; ab}_{\mathcal{O}; \rm Sea}$. 
The Fermi-surface contribution is given by, 
\begin{equation}
\alpha^{\gamma; ab}_{\mathcal{O}; \rm Surf} = -\sum_{n, \bm k} \tau \, \mathcal{J}^{\gamma;a}_{\mathcal{O}; nn} \bar{\varepsilon}_{n \bm k} v^b_{nn} \frac{\partial f_n}{\partial \bar{\varepsilon}_{n \bm k}}. \label{fermi_surf_contribution}
\end{equation}
Since $\alpha^{\gamma;ab}_{\mathcal{O};\rm Surf} \propto \tau$, this term represents the Drude-like extrinsic response.
To evaluate the Fermi-sea contribution, we consider the dilute-impurity limit characterized by \(\tau |\bar{\omega}_{nm}| \gg 1\). In this regime \cite{Harsh_2023}, 
\begin{equation}
g_{nm} = \left( \frac{1}{i \bar{\omega}_{nm}} + \frac{1}{\tau \bar{\omega}_{nm}^2} \right) \left( 1 + \frac{1}{\tau^2 \bar{\omega}_{nm}^2} \right)^{-1} \approx \frac{1}{i \bar{\omega}_{nm}}.
\end{equation}
Using this approximation together with the antisymmetry of $\Omega^{\gamma; ab}_{\mathcal{O}; mn}$ under exchange of band indices, the Fermi-sea contribution reduces to
\begin{align}
\alpha^{\gamma; ab}_{\mathcal{O};\rm Sea}=-\frac{1}{\hbar}\sum_{n, \bm k}\Big[\Omega^{\gamma; ab}_{\mathcal{O}; n}\, \sigma^3_{nn}\, \bar{\varepsilon}_{n \bm k}\, f_n-\hbar\epsilon_{abc}M^{\gamma; c}_{\mathcal{O}; n}\Big].\label{fermi_sea_contribution}
\end{align}
In contrast to the Fermi-surface term, the Fermi-sea contribution scales as $\tau^0$ and it is an intrinsic response. The single-band generalized Berry curvature is defined as \(\Omega^{\gamma; ab}_{\mathcal{O}; n}=\sum_{m \neq n}\Omega^{\gamma; ab}_{\mathcal{O}; nm},\)
where the summation runs over all other bands. While the Fermi-surface term involves contributions from all bands through $\partial f_n/\partial \bar{\varepsilon}_{n\bm k}$, the Fermi-sea term receives contributions only from magnon bands with $\bar{\varepsilon}_{n\bm k} \ge 0$.  Equations~(\ref{fermi_surf_contribution}) and~(\ref{fermi_sea_contribution}) are the main results of this work.

\section{Voltage Generated by EDM Accumulation}
In this section, we outline an electrical detection scheme for orbital transport. As discussed above, the orbital motion of magnons gives rise to an effective EDM. Consequently, an accumulation of EDMs at the sample boundaries is directly associated with an accumulation of MOMs.  
While the bulk theory assumes a spatially uniform current density $J^{\gamma;a}_{\mathcal{O}}$, finite-size effects prevent the transverse EDM current from flowing indefinitely, leading to edge accumulation and a spatially non-uniform polarization profile. This results in a spatially non-uniform polarization profile, which provides the basis for electrical detection \cite{Quang_2024}.

In the following, we consider the system to be two-dimensional (2D). To describe this effect, we relate the EDM current $J^{\gamma;a}_p$ to the electric polarization current density $J^{\gamma;a}_P$. Since electric polarization represents the EDM density, the corresponding polarization current is given by $J^{\gamma;a}_P = \frac{1}{d_z} J^{\gamma;a}_p$, where $d_z$ is the effective thickness of the system. Although the model is 2D, $d_z$ can be interpreted as the interlayer spacing in a van der Waals material for order-of-magnitude estimates. To account for boundary effects and relaxation processes, we employ the continuity equation for electric polarization~\cite{Bauer_2021, Bauer_2023, Adachi_2023},
\begin{equation} 
\partial_t P^{\gamma}(\bm r,t) + \nabla \cdot \bm J^{\gamma}_P(\bm r,t) = - \frac{P^{\gamma}(\bm r,t) - P^{\gamma}_0}{\tilde{\tau}}~.
\label{continuity_eq_electric_polarization}
\end{equation}
Here, $P^{\gamma}_0$ is the equilibrium polarization (arising from MOM and spin Berry curvature~\cite{Quang_2024}) and $\tilde{\tau}$ is the relaxation time of the non-equilibrium polarization. In the presence of both thermal field and edge accumulation, the polarization current along the $a$-direction is given by, 
\begin{equation}
J^{\gamma;a}_P(\bm r,t) = \frac{1}{d_z} J^{\gamma;a}_p - D \nabla_a P^{\gamma}(\bm r,t).
\label{polarization_current}
\end{equation}
Here, $D$ is the polarization diffusion coefficient and the second term accounts for diffusive backflow driven by spatial gradients of the polarization.  Solving Eqs.~(\ref{continuity_eq_electric_polarization}) and~(\ref{polarization_current}) gives the spatial profile of the non-equilibrium electric polarization $P^{\gamma}(\bm r,t)$ generated by the EDM current.

In the following, we consider a spatially uniform thermal gradient applied along the $\hat{y}$ direction. As the bulk EDM current density $J^{\gamma;a}_p$ is spatially uniform $ \nabla \cdot \mathbf{J}^{\gamma}_p = 0 $. As a result, in the steady state ($\partial_t P^\gamma = 0$), the continuity equation reduces to a Helmholtz-type equation for the polarization,
\begin{equation}
\nabla^2 P^{\gamma}(\bm r) = \frac{1}{\lambda^2} \left[ P^{\gamma}(\bm r) - P_0^{\gamma} \right], \label{Helmholtz_equation}
\end{equation}
where $\lambda = \sqrt{D \tilde{\tau}}$ defines the electric polarization diffusion length.
Since the thermal gradient is applied along $\hat{y}$, the induced EDM Nernst current flows transversely along $\hat{x}$. We therefore assume that the polarization varies only along the $x$-direction, reducing Eq.~(\ref{Helmholtz_equation}) to a one-dimensional form. The general solution is
\begin{equation}
P^{\gamma}(x) = P_0^{\gamma} + A \cosh\!\left(\frac{x}{\lambda}\right) +  B \sinh\!\left(\frac{x}{\lambda}\right),
\end{equation}
where $A$ and $B$ are constants determined by boundary conditions. We impose open-boundary conditions requiring that the polarization current vanishes at the sample edges,
\( J^{\gamma;x}_P(x=0)=0, \qquad J^{\gamma;x}_P(x=W)=0\), 
with $W$ the sample width along the $x$-direction. These conditions yield the spatial profile, 
\begin{equation}
P^{\gamma}(x) = P_0^{\gamma}-\frac{\lambda}{D d_z}J^{\gamma;x}_p\,\frac{\sinh\!\left(\dfrac{W-2x}{2\lambda}\right)}{\cosh\!\left(\dfrac{W}{2\lambda}\right)}~.
\label{electric_polarization}
\end{equation}

The spatially varying polarization generates an internal electric field along the transverse ($x$) direction. Within linear dielectric response, this field is given by
$ E^{x}(x) = \frac{P^{x}(x)}{\varepsilon_0 \chi} $, where $\varepsilon_0$ is the vacuum permittivity and $\chi$ is the electric susceptibility of the material.
From Eq.~(\ref{electric_polarization}), the polarization consists of a spatially uniform component $P_0^x$ and an antisymmetric part about $x=W/2$. As a result, the potential difference between the two edges ($x=0$ and $x=W$) vanishes due to symmetry. However, a finite voltage can be detected by measuring between the sample center and one edge. In this configuration, the measurable transverse voltage is
\begin{align}
V &= - \int_{W/2}^{W} E^x(x)\, dx \notag \\
& = - \frac{1}{\varepsilon_0 \chi} \left[ \frac{P_0^x W}{2} + \frac{\lambda^2}{D d_z} J^{x;x}_p \left( 1 - \sech\left(\frac{W}{2\lambda}\right) \right) \right]~.
\end{align} 
A measurable voltage arises only from the coexistence of both the magnon orbital Nernst effect and the magnon spin Nernst effect. The former generates the transverse EDM current $J^{x;x}_p$, while the latter contributes to the equilibrium polarization $P_0^x$. Their interplay produces the net detectable signal.

The equilibrium polarization $P_0^x$ was previously obtained in Ref.~\cite{Quang_2024} in the thermodynamic limit. Incorporating that result, we write
\begin{align}
P^x_0 &= \frac{g \mu_B}{\hbar W d_z c^2} \sum_{n, \bm k} \Big[ \Omega^{z;yx}_{s;n} \, k_B T \ln \big|e^{- \sigma^3_{nn} \bar{\varepsilon}_{n\bm k}/k_B T} -1 \big| \notag\\ &\qquad + 2 \hbar \sigma^3_{nn} s^z_{nn} L^z_{nn} f_n \Big]~. 
\end{align}
Here, $\Omega^{z;yx}_{s;n}$ denotes the spin Berry curvature obtained by replacing $\mathcal{O}$ with $s^z$ in Eq.~(\ref{generalized_QGT}). For a collinear antiferromagnet, $s^z_{nm} = \pm \delta_{nm}$ since the total spin operator commutes with the Hamiltonian~\cite{Neumann_2024}.

\begin{figure*}[t]
    \centering
    \includegraphics[width=0.85\linewidth]{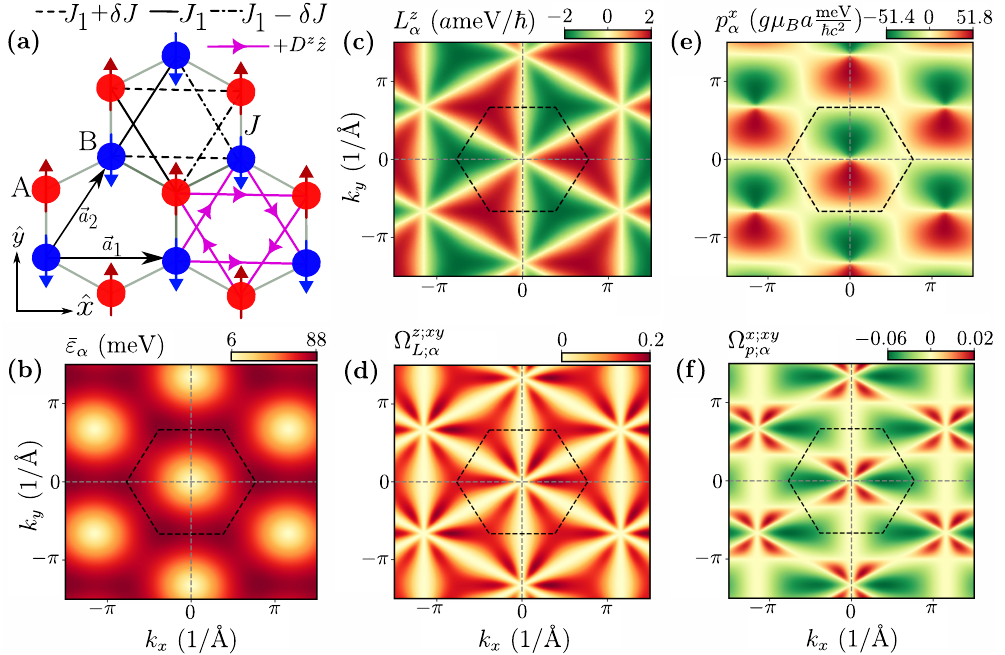}
    \caption{\justifying \textbf{Altermagnetic model and band-geometric properties of magnon bands.} 
    \textbf{(a)} Honeycomb lattice of the two-sublattice altermagnetic model. Red (spin-up) and blue (spin-down) arrows denote sublattices A and B, respectively. The NN isotropic exchange interaction is $J$, while anisotropic NNN exchange couplings are $J_1$, $J_1+\delta J$, and $J_1-\delta J$. The DMI vector for the NNN bonds indicated by pink arrows is $\bm D_{A(B)} = +D^z \hat{z}$. The real-space lattice vectors are $\bm a_1 = (\sqrt{3},0)a$ and $\bm a_2 = (\sqrt{3}/2,3/2)a$. 
    \textbf{(b)} Dispersion of the $\alpha$ magnon band, with the hexagonal Brillouin zone boundary shown by the dotted line.
    \textbf{(c)} and \textbf{(d)} Momentum-space distributions of the MOM and the corresponding OBC of the $\alpha$ mode.
    \textbf{(e)} and \textbf{(f)} Momentum-space distributions of the EDM and the corresponding DBC of the $\alpha$ mode.
    The parameters used are $\{J, J_1, \delta J, K, D^z\} = \{12.4, -5.48, 0.8, -0.3, 0.3\}\,\mathrm{meV}$ with spin $S=1$.
    \label{fig:model_BGQs}}
\end{figure*}
\section{Thermoelectric orbital responses in a Hexagonal Altermagnet}
\label{sea:model_calculation}
To apply the general transport formalism developed above for quantitative estimates, we consider a collinear altermagnet on a hexagonal lattice \cite{Sankar_2025}. The spin Hamiltonian is given by 
\begin{align}
    \mathcal{H} &= J \sum_{\langle i\in A, j\in B \rangle} \bm S_i \cdot \bm S_j + K \sum_i (S_i^z)^{2} \nn \\
    &\quad +\frac{1}{2} \sum_{m =1}^6 \sum_{i \in A} [J_{mA} \bm S_i \cdot \bm S_{i + \delta_m}+\bm D_A \cdot (\bm S_i \times \bm S_{i + \delta_m})] \nn \\
    &\quad +\frac{1}{2} \sum_{m =1}^6 \sum_{i \in B} [J_{mB} \bm S_i \cdot \bm S_{i + \delta'_m}+\bm D_B \cdot (\bm S_i \times \bm S_{i + \delta'_m})]~.
    \label{spin_hamiltonian}
\end{align}
The first term describes the antiferromagnetic nearest-neighbor (NN) exchange interaction ($J>0$) between spins on sublattices $A$ and $B$. The second term represents a uniaxial anisotropy ($K<0$) that stabilizes the N\'eel order along the $z$-axis.  The altermagnetic character arises from anisotropic next-nearest-neighbor (NNN) exchange interactions within each sublattice. For sublattice $A$ ($B$), the anisotropic couplings are denoted by $J_{mA}$ ($J_{mB}$), connecting spins at site $i$ and $i+\bm\delta_m$ ($i+\bm\delta'_m$). The NNN vectors for sublattice $A$ are
\( 
\bm\delta_1 = \left(-\frac{\sqrt{3}}{2},\frac{3}{2}\right)a,\quad \bm\delta_2 = \left(-\frac{\sqrt{3}}{2},-\frac{3}{2}\right)a,\quad \bm\delta_3 = (\sqrt{3},0)a,
\)
with $\bm\delta_{4,5,6} = -\bm\delta_{1,2,3}$. For sublattice $B$, the NNN vectors satisfy $\bm\delta'_m = -\bm\delta_m$. The Hamiltonian also includes symmetry-allowed Dzyaloshinskii–Moriya interactions (DMI) between NNN sites. The DMI enhances asymmetric magnon band splitting and plays a crucial role in generating Nernst-type responses of EDMs \cite{Quang_2024}. For sublattice $A$, the DMI vector is $\bm D_A = +D^z \hat{z}$ for clockwise-oriented bonds along $\bm\delta_{1,2,3}$ and $\bm D_A = -D^z \hat{z}$ along $\bm\delta_{4,5,6}$. For sublattice $B$, the same pattern applies with $\bm\delta_m \rightarrow \bm\delta'_m$. Following Ref.~\cite{Sankar_2025}, we choose the anisotropic NNN exchange couplings as $ J_{1A} = J_{4A} = J_1, ~~J_{2A} = J_{5A} = J_1 - \delta J, ~~ J_{3A} = J_{6A} = J_1 + \delta J$. Additionally, we choose $J_{1B} = J_{4B} = J_1 - \delta J, ~~ J_{2B} = J_{5B} = J_1, ~~ J_{3B} = J_{6B} = J_1 + \delta J$, as illustrated in Fig.~\ref{fig:model_BGQs}(a). This pattern relates the two sublattices through the composite symmetry operation [$C_2 || M_y$], where $C_2$ is a twofold rotation in spin space and $M_y$ is a mirror reflection perpendicular to the $y-$axis.

\subsection{Magnon dispersion and band geometry}
In this work, we focus on the low-magnon-density regime and neglect magnon–magnon interactions. This approximation is justified when the temperature of interest is well below the N\'eel temperature, such that the antiferromagnetic order remains robust and the linear spin-wave theory is valid. To this end, we use the Holstein–Primakoff (HP) transformation for a bipartite antiferromagnet \cite{Holstein_1940, Kubo_1952}. The spin operators on sublattices $A$ and $B$ are expressed as
\begin{align}
S_i^{+} &\approx \sqrt{2S}\, a_i, ~  S_i^{-} \approx \sqrt{2S}\, a_i^{\dagger}, ~~  S_i^{z} = S - a_i^{\dagger} a_i,~ (i \in A), \notag \\ S_j^{+} &\approx \sqrt{2S}\, b_j^{\dagger}, ~~  S_j^{-} \approx \sqrt{2S}\, b_j,  \quad  S_j^{z} = -S + b_j^{\dagger} b_j, ~ (j \in B).
\end{align}
Here, $S_i^{\pm} = (S_i^x \pm i S_i^y)/2$ are the spin raising and lowering operators at lattice site $i$. The operators $a_i^\dagger$ ($b_i^\dagger$) are magnon creation operator for sublattice $A$ ($B$) at site $i$, while $a_i$ ($b_i$) annihilate them. These bosonic operators satisfy the canonical commutation relations
$[a_i, a_j^\dagger] = [b_i, b_j^\dagger] = \delta_{ij}$, and all other commutators vanish. 

Substituting the HP transformation into Eq.~\eqref{spin_hamiltonian} and retaining terms up to quadratic order in bosonic operators yields the real-space magnon Hamiltonian. We transform it to momentum space via the Fourier transformation, 
\begin{align}
\begin{pmatrix} a_i \\ b_i \end{pmatrix} =
\frac{1}{\sqrt{N}} \sum_{\bm k} e^{i \bm k \cdot \bm r_i} \begin{pmatrix} a_{\bm k} \\ b_{\bm k} \end{pmatrix},
\end{align}
with $N$ the number of unit cells. This gives a quadratic Hamiltonian of the form,
$\mathcal{H} = \frac{1}{2} \sum_{\bm k} \Psi_{\bm k}^{\dagger} \mathcal{H}_{\bm k} \Psi_{\bm k}$, 
where the Nambu basis is
\( \Psi_{\bm k} = \left( a_{\bm k}, b_{\bm k}, a_{-\bm k}^{\dagger}, b_{-\bm k}^{\dagger} \right)^T \). 
The momentum-space bosonic operators satisfy
\( [a_{\bm k}, a_{\bm k'}^{\dagger}] = [b_{\bm k}, b_{\bm k'}^{\dagger}] = \delta_{\bm k \bm k'}\), with all other commutators equal to zero. The Bogoliubov–de Gennes magnon Hamiltonian is 
\begin{align}
\mathcal{H}_{\bm k} =
\begin{pmatrix}
J' + f_A^- & 0 & 0 & JS f^* \\
0 & J' + f_B^+ & JS f & 0 \\
0 & JS f^* & J' + f_A^+ & 0 \\
JS f & 0 & 0 & J' + f_B^-
\end{pmatrix},
\end{align}
where $J' = 3JS - 2KS$ and $f_{A(B)}^{\pm}(\bm k) = f_{A(B)}(\bm k) \pm \eta(\bm k)$. The structure factors are given by
\begin{align}
f(\bm k) &= 2 e^{i a_0 k_y/2} \cos\!\left(\frac{\sqrt{3} a_0 k_x}{2}\right) + e^{-i a_0 k_y}, \\
f_A(\bm k) &= 2S \sum_{m=1}^{3} J_{mA} \left[ \cos(\bm k \cdot \bm\delta_m) - 1 \right], \\
f_B(\bm k) &= 2S \sum_{m=1}^{3} J_{mB} \left[ \cos(\bm k \cdot \bm\delta_m) - 1 \right], \\
\eta(\bm k) &= 2 D^z S \sum_{m=1}^{3} \sin(\bm k \cdot \bm\delta_m). 
\end{align}

As $\mathcal{H}_{\bm k}$ is written in a bosonic Nambu basis, a conventional unitary diagonalization does not preserve the canonical commutation relations. Instead, the Hamiltonian has to be diagonalized using a para-unitary transformation. We introduce a transformation matrix $P$ such that
$ \Phi_{\bm k} = P^{-1} \Psi_{\bm k} = (\alpha_{\bm k}, \beta_{\bm k}, \alpha_{-\bm k}^{\dagger}, \beta_{-\bm k}^{\dagger})^T$. This diagonalizes the Hamiltonian to be, 
$ \mathcal{H} = \frac{1}{2} \sum_{\bm k} \Psi_{\bm k}^\dagger \mathcal{H}_{\bm k} \Psi_{\bm k} = \frac{1}{2} \sum_{\bm k} 
\Phi_{\bm k}^\dagger \mathcal{E}_{\bm k} \Phi_{\bm k}. $
Here, $ \mathcal{E}_{\bm k} = \mathrm{diag} \left( \varepsilon_{1,\bm k}, \varepsilon_{2,\bm k}, \varepsilon_{1,-\bm k}, \varepsilon_{2,-\bm k} \right)$, 
contains the magnon band energies. To preserve the bosonic commutation relations, the transformation matrix must satisfy the para-unitary condition
$ P \sigma^3 P^\dagger = P^\dagger \sigma^3 P = \sigma^3$.  Combining this condition with the diagonalization equation, 
\( \mathcal{E}_{\bm k} = P^\dagger \mathcal{H}_{\bm k} P \),  we obtain $ P^{-1} \sigma^3 \mathcal{H}_{\bm k} P = \sigma^3 \mathcal{E}_{\bm k} \equiv \bar{\mathcal{E}}_{\bm k} $. Here, $ \bar{\mathcal{E}}_{\bm k} = \mathrm{diag} \left( \varepsilon_{1,\bm k}, \varepsilon_{2,\bm k}, -\varepsilon_{1,-\bm k}, -\varepsilon_{2,-\bm k} \right)$.  
This implies that the magnon spectra and eigenstates can be obtained by solving the following eigenvalue equation,  $ \bar{\mathcal{H}}_{\bm k}$ \cite{Gyungchoon_2024, Daehyeon_2025, Quang_2024, Li_2020}: $ \bar{\mathcal{H}}_{\bm k} \ket{u^R_{n \bm k}} =\bar{\varepsilon}_{n \bm k} \ket{u_{n \bm k}^R} $ and $ \bra{u_{n \bm k}^L} \bar{\mathcal{H}}_{\bm k} = \bar{\varepsilon}_{n \bm k} \bra{u_{n \bm k}^L} $. Here, the right-eigenvector $ \ket{u_{n \bm k}^R} = \ket{u_{n \bm k}} $ is the column vector of $P$ and left-eigenvector is $ \bra{u_{n \bm k}^L} = \bra{u_{n \bm k}} \sigma^3 $ with the generalized orthogonal condition $ \braket{u_{n \bm k}^L | u_{m \bm k}} = \delta_{nm} $.

Having obtained the eigenvalues and eigenvectors, we now analyze the magnon dispersion and associated band-geometric quantities.  In Fig.~(\ref{fig:model_BGQs}\textcolor{blue}{b}) we present the dispersion of the first ($\alpha-$mode) magnon band. The $\bm k-$space distribution of MOM and OBC of the corresponding magnon band are displayed in Figs.~(\ref{fig:model_BGQs}\textcolor{blue}{c}) and (\ref{fig:model_BGQs}\textcolor{blue}{d}), respectively. Due to interband coherence, these band geometric quantities show enhanced values near the band crossing. Figs.~(\ref{fig:model_BGQs}\textcolor{blue}{e}) and (\ref{fig:model_BGQs}\textcolor{blue}{f}) show the EDM and DBC distribution over $\bm k-$space for $\alpha-$mode.   
\subsection{Nernst and Seebeck responses of MOM and EDM}
We now present the numerical results for the Nernst- and Seebeck-type transport coefficients of the MOM and the associated EDM. For concreteness, we consider a uniform temperature gradient applied along the $\hat{y}$ direction, corresponding to the thermal field  
\( \bm E_T = -\frac{\nabla_y T}{T} \hat{y} \). Since the model is 2D with N\'eel order along $\hat{z}$, the magnon orbital moment is oriented out of plane. In contrast, the EDM $\bm p \propto (\bm v \times \bm m - \bm m \times \bm v)$ lies in the 2D plane and can have only $x$- and $y$-components. In the following, we focus on the $x$-component of the EDM, which is relevant for the transverse voltage discussed in the next section.
\begin{figure}
    \centering
    \includegraphics[width=1\linewidth]{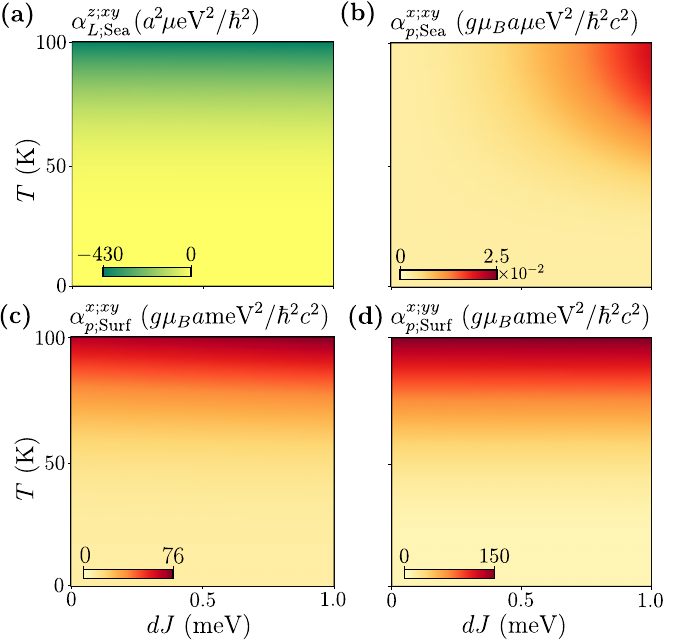}
    \caption{\justifying \textbf{Variation of the response tensors} in the $(\delta J, T)$ parameter space  for a thermal field $\bm E_T = -(\nabla_y T/T)\hat{y}$. 
    \textbf{(a)} Fermi-sea contribution to the MOM Nernst effect of the MOM polarized along $\hat{z}$. 
    \textbf{(b)} and \textbf{(c)} Fermi-sea and Fermi-surface contributions, respectively, to the EDM Nernst effect of the EDM oriented along $\hat{x}$-axis.  
    \textbf{(d)} Fermi-surface contribution to the EDM Seebeck effect of the EDM with the same orientation. All parameters are identical to those used in Fig.~\ref{fig:model_BGQs}. 
    \label{fig:Response_T_dJ}}
\end{figure}
\begin{figure}[t]
    \centering
    \includegraphics[width=1.0\linewidth]{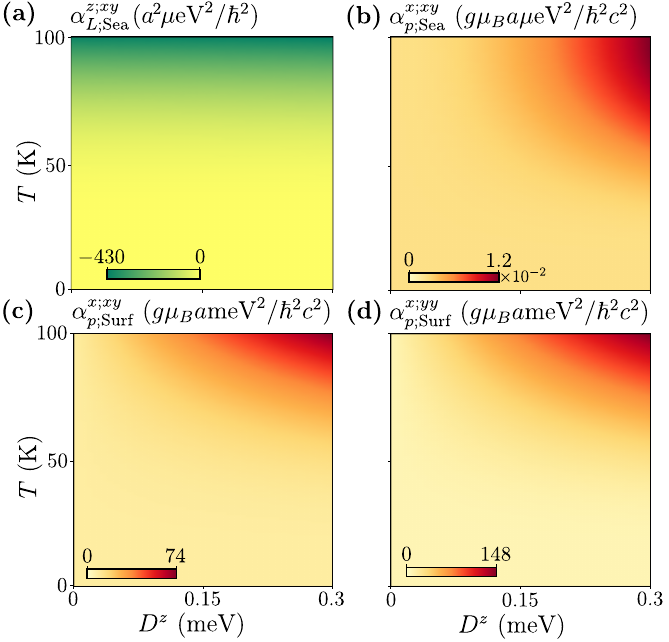}
    \caption{\justifying \textbf{DMI and temperature dependence of transport coefficients} for thermal field $\bm E_T = - (\nabla_y T/T)\hat{y}$. 
    \textbf{(a)} Fermi-sea contribution to the MOM Nernst effect with moment oriented along $\hat{z}$. 
    \textbf{(b,c)} Fermi-sea and Fermi-surface contributions to the EDM Nernst response for dipole oriented along $\hat{x}$. 
    \textbf{(d)} Fermi-surface contribution to the corresponding EDM Seebeck response. 
    All other parameters are the same as in Fig.~\ref{fig:model_BGQs}.
    \label{fig:Response_T_Dz}} 
\end{figure}
Figures~\ref{fig:Response_T_dJ}(a) and \ref{fig:Response_T_Dz}(a) show the Fermi-sea contribution to the MOM Nernst coefficient in the $(\delta J, T)$ and $(D^z, T)$ parameter spaces, respectively. This response originates from the OBC $\Omega^{z;xy}_{L;n}$. The OBC is an even function over the Brillouin zone and it arises intrinsically from the broken inversion symmetry of the honeycomb lattice. Importantly, its $\bm k$-space distribution is nearly insensitive to both $\delta J$ and $D^z$. Consequently, the Fermi-sea MOM Nernst response is essentially independent of these parameters. In contrast, the response increases monotonically with temperature owing to the increasing thermal population of magnon states. 

The Fermi-sea contribution to the EDM Nernst effect is shown in Figs.~\ref{fig:Response_T_dJ}(b) and \ref{fig:Response_T_Dz}(b). Unlike the OBC, the DBC originates from anisotropic NNN exchange interactions and therefore vanishes when $\delta J = 0$. In the absence of DMI, the $\bm k$-space distributions of the DBC for the $\alpha$ and $\beta$ bands are equal in magnitude and opposite in sign, leading to an exact cancellation of the Fermi-sea contribution. The inclusion of DMI breaks this cancellation by lifting the symmetry between the two bands, resulting in a finite intrinsic EDM Nernst response. As evident from Figs.~\ref{fig:Response_T_dJ}(b) and \ref{fig:Response_T_Dz}(b), the Fermi-sea contribution increases with both $\delta J$ and $D^z$, and vanishes when either of them is set to zero. 

The Fermi-surface contribution can be understood from the $\bm k$-space distribution of the EDM shown in Fig.~\ref{fig:model_BGQs}(e). In the absence of DMI, the dipole moment $p^x_\alpha(k_x,k_y)$ is odd in $k_y$, i.e.,  \( p^x_\alpha(k_x,-k_y) = -p^x_\alpha(k_x,k_y) \). This leads to a vanishing Brillouin-zone integral and therefore zero Fermi-surface response. The presence of DMI breaks this antisymmetry and allows a finite contribution. This behavior is reflected in Figs.~\ref{fig:Response_T_dJ}(c) and \ref{fig:Response_T_Dz}(c), where the Fermi-surface EDM Nernst coefficient $\alpha^{x;xy}_{p;\rm Surf}$ is shown. These coefficient weakly depends on $\delta J$ but increase significantly with DMI strength. Figures~\ref{fig:Response_T_dJ}(d) and \ref{fig:Response_T_Dz}(d) display the Fermi-surface contribution to the EDM Seebeck response. Similar to the Nernst case, the Seebeck coefficient grows with increasing $D^z$.  Overall, both intrinsic (Fermi-sea) and extrinsic (Fermi-surface) EDM responses are strongly controlled by the interplay between anisotropic exchange ($\delta J$) and DMI ($D^z$), while their temperature dependence reflects thermally activated magnon transport.
\begin{figure}[t]
    \centering
\includegraphics[width=1.02\linewidth]{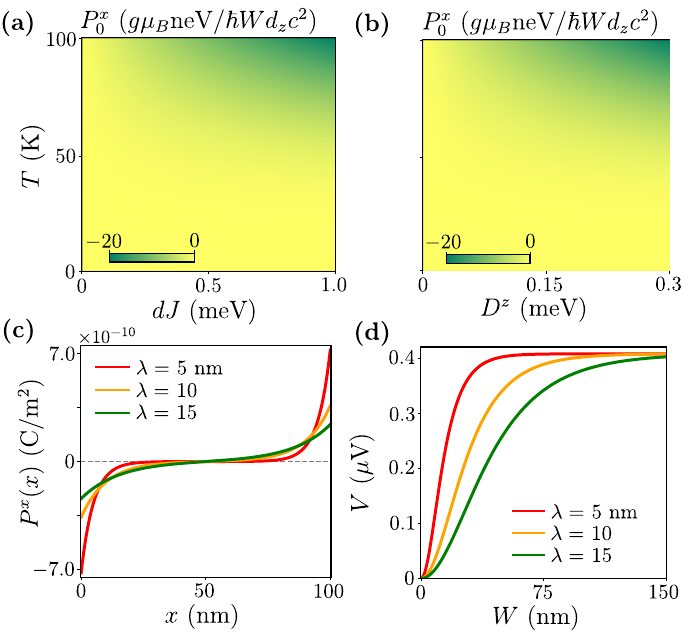}
    \caption{\justifying \textbf{Equilibrium polarization and induced transverse voltage.} 
    \textbf{(a,b)} Equilibrium polarization $P_0^x$ as functions of $(\delta J, T)$ and $(D^z, T)$, respectively. 
    \textbf{(c)} Spatial profile of electric polarization $P^x(x)$ for a transverse temperature gradient. 
    \textbf{(d)} Transverse electric voltage as a function of sample width $W$. 
    Parameter values: $D^z = 0.3$ meV for \textbf{(a,c,d)}, $\delta J = 1.0$ meV for \textbf{(b,c,d)}, $T = 100$ K, $\tilde{\tau} = 1$ ns, $a = 1~\text{\AA}$, and $\chi = 1$ for \textbf{(c,d)}. 
    All remaining parameters are the same as in Fig.~\ref{fig:model_BGQs}.
    \label{fig:Polarization_voltage}}
\end{figure}
\subsection{Electrical Detection of Magnon Orbital Transport}
We now combine the EDM Nernst response with the equilibrium polarization to estimate the measurable electric voltage. 
Figures~\ref{fig:Polarization_voltage}(a) and \ref{fig:Polarization_voltage}(b) show the equilibrium polarization $P_0^x$ in the $(\delta J,T)$ and $(D^z,T)$ parameter spaces, respectively. Consistent with the intrinsic EDM Nernst response, the equilibrium polarization increases with both the anisotropic NNN exchange strength $\delta J$ and the DMI amplitude $D^z$. Its magnitude also grows with temperature due to enhanced thermal population of magnon states.
The spatial profile of the non-equilibrium polarization $P^x(x)$, shown in Fig.~\ref{fig:Polarization_voltage}(c), depends sensitively on the diffusion length $\lambda = \sqrt{D\tilde{\tau}}$. For shorter diffusion lengths, the polarization accumulates more strongly near the sample edges, whereas for larger $\lambda$ it spreads more uniformly across the sample width. 

Using this polarization profile, we estimate the transverse electric voltage. Although the model is strictly 2D, we adopt an effective monolayer thickness $d_z \approx 3\,\text{\AA}$ for order-of-magnitude estimates, consistent with typical van der Waals materials \cite{Sankar_2025, Liu_2021, Zhang_2023, Jinghao_2025}. We consider a representative temperature gradient of $10~\text{K}/\mu\text{m}$ at $T=100~\text{K}$ \cite{Kryder_2008}. This corresponds to a thermal field strength $\nabla_y T/T = 10^5~\text{m}^{-1}$. Figure~\ref{fig:Polarization_voltage}(d) displays the resulting voltage as a function of the sample width $W$ for different diffusion lengths. When $W$ is comparable to $\lambda$, the voltage increases with $W$ because the polarization remains spatially non-uniform across the sample. In contrast, for $W \gg \lambda$, the voltage saturates and becomes independent of the sample width. In this regime, the contribution from edge accumulation becomes negligible and the signal is dominated by the uniform equilibrium polarization. For the chosen parameters, the maximum voltage is estimated to be approximately $0.4~\mu\text{V}$, which lies well within the sensitivity range of modern experimental detection techniques.
\section{Conclusion}
In this work, we developed a density-matrix-based quantum-kinetic framework for the thermal transport of magnon orbital moments and the associated orbital-motion induced electric dipole moments. Our formulation captures both the scattering-time dependent Drude-like extrinsic response and the intrinsic response governed by band geometry. In particular, we show that the intrinsic MOM and EDM Nernst effects originate from the orbital Berry curvature and dipolar Berry curvature, respectively. We further proposed an electrical detection scheme for magnon orbital transport based on the accumulation of EDMs at sample boundaries. By incorporating finite-size effects through the continuity equation for electric polarization, we demonstrated how the interplay between non-equilibrium EDM currents and equilibrium polarization generates a measurable voltage signal. 

To illustrate the theory and assess experimental feasibility, we applied the formalism to a hexagonal insulating altermagnet with anisotropic next-nearest-neighbor exchange interactions and Dzyaloshinskii–Moriya interaction. While the MOM Nernst response arises purely from the Fermi-sea contribution, the EDM Nernst and Seebeck responses contain both Fermi-sea and Fermi-surface components. For experimentally accessible thermal gradients, the resulting transverse voltage is estimated to be approximately $0.4~\mu$V, well within current detection capabilities. Overall, our results establish a direct electrical probe of magnon orbital transport and highlight insulating altermagnets as promising platforms for low-dissipation orbitronics phenomena.

\section*{Acknowledgments}
S.S. acknowledges the Indian Institute of Technology Kanpur for the Ph.D. fellowship. A.A. acknowledges funding from the Core Research Grant by the
Anusandhan National Research Foundation (ANRF, Sanction No. CRG/2023/007003), Department of Science and
Technology, India. A.A. acknowledges the high-performance
computing facility at IIT Kanpur, including HPC 2013, and
Param Sanganak.

\appendix
\section{Density matrix calculation}\label{details_of_first_order_DM}
In this section, we present the detailed calculations of the density matrix for a $ d.c.$ thermal field. Using the perturbative series expansion of $\rho(\bm k)$ in Eq. \eqref{quantum_lioiville_equation} along with the steady state condition, the quantum Liouville equation for the first-order density matrix becomes
\begin{equation}
    \frac{i}{\hbar} [ \bar{\mathcal{H}}_{\bm k}, \rho^{(1)} ]_{nm} + \frac{\rho^{(1)}_{nm}}{\tau} = \mathcal{D}_T[\rho^{(0)}]_{nm}~.\label{QKE_app}
\end{equation}
The interband-matrix elements of $ [ \bar{\mathcal{H}}_{\bm k}, \rho^{(1)} ]_{nm} $ are
\begin{align*}
     [ \bar{\mathcal{H}}_{\bm k}, \rho^{(1)} ]_{nm} &= \Big[ \bra{u_n} \sigma^3 \bar{\mathcal{H}}_{\bm k} \rho^{(1)} \ket{u_m} - \bra{u_n} \sigma^3 \rho^{(1)} \bar{\mathcal{H}}_{\bm k}  \ket{u_m} \Big]\\
    &= \sum_q  \sigma^3_{qq} \Big[ \bra{u_n} \sigma^3 \bar{\mathcal{H}}_{\bm k} \ket{u_q} \bra{u_q} \sigma^3 \rho^{(1)} \ket{u_m} \\
    &\quad - \bra{u_n} \sigma^3 \rho^{(1)} \ket{u_q} \bra{u_q} \sigma^3 \bar{\mathcal{H}}_{\bm k}  \ket{u_m} \Big] \\
    &= \hbar \bar{\omega}_{nm} \rho^{(1)}_{nm~.}
\end{align*}
Here, in the second line and third line, we have used the identity matrix $\mathcal{I} = \sum_q \sigma^3_{qq} \ket{u_q} \bra{u_q} \sigma^3 $ and orthonormality condition $ \sigma^3_{qq} \sigma^3_{qn} = \delta_{qn} $, respectively. Also, for simplicity, we have dropped the momentum index of state $\ket{u_{n \bm k}}$. The right hand side of Eq. (\ref{QKE_app}), $ \mathcal{D}[\rho^{(0)}]_{nm} $ simplifies to
\begin{align*}
    &=  \frac{-E_T^a}{2 \hbar} \sum_q \sigma^3_{qq}  \Big[ \bra{u_n} \sigma^3 \bar{\mathcal{H}}_{\bm k} \ket{u_q} \bra{u_q} \sigma^3 \partial_{k_a} \rho^{(0)} \ket{u_m} \\ 
    &\quad + \bra{u_n} \sigma^3 \partial_{k_a} \rho^{(0)}  \ket{u_q} \bra{u_q} \sigma^3 \bar{\mathcal{H}}_{\bm k} \ket{u_m} \\ 
    &\quad - i \Big( \mathcal{R}^a_{nq} \{ \bar{\mathcal{H}}_{\bm k}, \rho^{(0)} \}_{qm} - \mathcal{R}^a_{qm} \{ \bar{\mathcal{H}}_{\bm k}, \rho^{(0)} \}_{nq}  \Big) \Big] \\
    &= - \frac{E_T^a}{\hbar} [ \bar{\varepsilon}_{n \bm k} \partial_{k_a} f_n \delta_{nm} + i \mathcal{R}^a_{nm} \xi_{nm} ]~.
\end{align*}
Here, we have defined $ \xi_{nm} = ( \sigma^3_{nn} \bar{\varepsilon}_{n \bm k} f_n - \sigma^3_{mm} \bar{\varepsilon}_{m \bm k} f_m ) $. Inserting the matrix elements of $ [\bar{\mathcal{H}}_{n \bm k}, \rho^{(1)}] $ and $ \mathcal{D}_T[\rho^{(0)}] $ into Eq. (\ref{QKE_app}), we get the first order density matrix to be
\begin{equation}
    \rho^{(1)}_{nm} = - \frac{E_T^a}{\hbar} [ \tau \bar{\varepsilon}_{n \bm k} \partial_a f_n \delta_{nm} + i g_{nm} \mathcal{R}^a_{nm} \xi_{nm} ]~.\label{first_order_DM_app}
\end{equation}

\section{Magnon Hamiltonian}\label{magnon_hamiltonian_appendix}
Here, we explicitly show the derivation of the magnon Hamiltonian starting from the spin Hamiltonian for an antiferromagnet. The spin Hamiltonian, as presented in the main text, is given by
\begin{align}
    \mathcal{H} &= J \sum_{\langle i\in A, j\in B \rangle} \bm S_i \cdot \bm S_j + K \sum_i (S_i^z)^{2} \nn \\
    &\quad +\frac{1}{2} \sum_{m =1}^6 \sum_{i \in A} [J_{mA} \bm S_i \cdot \bm S_{i + \delta_m}+\bm D_A \cdot (\bm S_i \times \bm S_{i + \delta_m})] \nn \\
    &\quad +\frac{1}{2} \sum_{m =1}^6 \sum_{i \in B} [J_{mB} \bm S_i \cdot \bm S_{i + \delta'_m}+\bm D_B \cdot (\bm S_i \times \bm S_{i + \delta'_m})].
\end{align}
 Here, the first term describes NN antiferromagnetic exchange (\( J > 0 \)) between sublattices \( A \) and \( B \), and the uniaxial anisotropy term (\( K < 0 \)) ensures collinear ordering. The second term represents NNN interactions within sublattice \( A \), with altermagnetic exchanges \( J_{mA} \). The third term describes NNN interactions within sublattice \( B \), with exchanges \( J_{mB} \). The NNN position vectors for sublattice \( A \) are \( \delta_1 = \left( -\sqrt{3}a/2, 3a/2 \right) \),  \(\delta_2 = \left( -\sqrt{3}a/2, -3a/2 \right) \), and \(\delta_3 = \left( \sqrt{3}a, 0 \right)\), and  \(\delta_{4,5,6} = -\delta_{1,2,3}\). For sublattice B the NNN positions vectors are opposite of those for A i.e., $\delta^{\prime}_m = - \delta_m$.


%
 For simplicity, we simplify the spin Hamiltonian term by term as follows. By applying the HP and Fourier transformations, the NN interaction is given by
\begin{align}
    \mathcal{H}_{NN} &= JS \sum_{\langle i\in A, j \in B \rangle} ( a_i^{\dagger} a_i + b_j^{\dagger} b_j + a_j b_j + a_i^{\dagger} b_j^{\dagger} ) \notag\\
    &= JS \sum_{\bm k} (3 a_{\bm k}^{\dagger} a_{\bm k} + 3 b_{\bm k}^{\dagger} b_{\bm k} + f(\bm k)^* a_{\bm k} b_{-\bm k} + f(\bm k) a_{\bm k}^{\dagger} b_{-\bm k}^{\dagger} ).
\end{align}
In the second line we used the identity $\delta(\bm k - \bm k^{\prime} ) = (1/N) \sum_i e^{i(\bm k - \bm k^{\prime}) \cdot \bm r_i}$. The structure factor $f(\bm k)$ for NN is given by
\begin{equation*}
    f(\bm k) = \sum_{j = 1}^3 e^{i \bm k \cdot \bm d_j} = 2 e^{iak_y/2} \text{cos}( \sqrt{3} a k_x/2) + e^{-iak_y}~,
\end{equation*}
where we use the NN position vectors: $ \bm d_1 = a(\sqrt{3}/2, 1/2),~ \bm d_2 = a( -\sqrt{3}/2, 1/2 ) $ and $ \bm d_3 = a(0, -1) $. The NNN interaction for sublattice A is given by
\begin{widetext}
\begin{align}
    \mathcal{H}_{NNN, A} &= S \sum_{m = 1}^3 \sum_{i \in A} [ J_{mA} (a_{i} a_{i + \delta_m}^{\dagger} + a_i^{\dagger} a_{i + \delta_m} )  - J_{mA} (a_i a_i^{\dagger} + a_{i + \delta_m}^{\dagger} a_{i +\delta_m}) - iD^z (a_i a_{i +\bm \delta_m}^{\dagger} - a_i^{\dagger} a_{i + \bm \delta_m}) ] \notag\\
    &= \frac{1}{2} \sum_{\bm k} [f^-_A(\bm k)  a_{\bm k}^{\dagger} a_{\bm k} + f_A^+ (\bm k) a_{-\bm k} a_{-\bm k}^{\dagger} ]~.
\end{align}
Similarly, the NNN interaction for sublattice B is given by
\begin{equation}
    \mathcal{H}_{NNN, B} = \frac{1}{2} \sum_{\bm k} [f^+_B(\bm k)  b_{\bm k}^{\dagger} b_{\bm k} + f_B^{-} (\bm k) b_{-\bm k} b_{-\bm k}^{\dagger} ]~.
\end{equation}
The structure factor NNN interaction for sublattices A and B are respectively given by
\begin{align}
    f_A(\bm k) &= 2S \sum_m  J_{mA}  \{ \text{cos}(\bm k \cdot \bm \delta_m ) - 1 \} ,\nn \\
    f_B(\bm k) &= 2S \sum_m  J_{mB} \{ \text{cos} (\bm k \cdot \bm \delta_m) - 1 \} ,\nn\\
    \eta(\bm k) &= 2D^z S \sum_{m = 1}^3 \sin(\bm k \cdot \bm \delta_m)~.
\end{align}
\end{widetext}
Finally, the anisotropic energy term is given by
\begin{equation}
    \mathcal{H}_{\text{aniso}} = - KS \sum_{\bm k} ( a_{\bm k}^{\dagger} a_{\bm k} + b_{\bm k}^{\dagger} b_{\bm k}~. )
\end{equation}
Combining all the interactions term, the total magnon Hamiltonian is given by: $H = \frac{1}{2} \sum_{\bm k} \Psi_{\bm k}^{\dagger} \mathcal{H}_{\bm k} \Psi_{\bm k} $, where the momentum space magnon Hamiltonian is
\be
\mathcal{H}_{\bm k} = \begin{pmatrix}
    J' + f^-_A & 0 & 0 & JSf^* \\
    0 & J' + f^+_B  & JSf & 0 \\
    0 & JSf^* & J' + f^+_A  & 0 \\
    JSf & 0 & 0 & J' + f^-_B 
\end{pmatrix}.\label{magnon_Hamiltonian}
\ee
Here, $ \Psi_{\bm k} = [ a_{\bm k},~ b_{\bm k}, ~ a^{\dagger}_{-\bm k},~ b^{\dagger}_{-\bm k} ]^T $, $J' = (3JS - 2KS )$ and $ f^{\pm}_{A(B)}  = f_{A(B)} (\bm k) \pm \eta (\bm k) $.

\section{Interband matrix elements of magnon orbital and electric dipole moment}
In this section, we give a comprehensive expression of the matrix element of the MOM. As defined in the main text, $a$-component MOM is $ \hat{L}^a = (1/4) \epsilon_{abc} (\hat{r}^b \hat{v}^c - \hat{v}^b \hat{r}^c) $. The action of position operator on periodic part of Bloch wavefunction, $\ket{u_{n\bm k}}$ is given by $ \hat{r}^c \ket{u_{n \bm k}} = i \ket{\partial_{k_c} u_{n \bm k}} $ and the velocity operator is $ \hat{v}^b = (-i/\hbar) [ \hat{r}^c, \mathcal{H}_{\bm k}] $. Using these identity we can write
\begin{align}
    L^a_{np} &= - \frac{i}{4 \hbar} \epsilon_{abc} \bra{u_n} \hat{r}^b \hat{r}^c \mathcal{H}_{\bm k} - 2 \hat{r}^b \mathcal{H}_{\bm k} \hat{r}^c + \mathcal{H}_{\bm k} \hat{r}^b \hat{r}^c \ket{u_{p}} \notag\\
    &= - \frac{i}{4 \hbar} \epsilon_{abc}  \bra{u_n} (\bar{\varepsilon}_n + \bar{{\varepsilon}}_p )  \hat{r}^b \sigma^3 \hat{r}^c - 2 \hat{r}^b \mathcal{H}_{\bm k} \hat{r}^c \ket{u_p} \notag\\
    &= \frac{i}{4 \hbar} \bra{\partial_{\bm k} u_n} \times [( \bar{\varepsilon}_n + \bar{\varepsilon}_p ) \sigma^3 - 2 \mathcal{H}_{\bm k}] \ket{\partial_{\bm k} u_p}~.\label{L_a_np}
\end{align}
In the simplification of the second line, we have that $ [\hat{r}^b, \sigma^3] = 0 $. To simplify the above expression further, we need to evaluate the momentum derivative terms of the Bloch functions. To do this, we use the relation between the interband Berry connection and the interband velocity element, which is
\begin{align}
    \mathcal{R}^a_{np} &= i \sigma^3_{nn} \bra{u_n} \sigma^3 \ket{\partial_{k_a} u_p} = - \frac{i \sigma^3_{nn}}{\bar{\omega}_{np}} v^a_{np} ~.\notag
\end{align}
Multiplying both side of the above equation by $\ket{u_n}$ and taking sum over $n$, we get
\begin{align}
\Big( \sum_{n \neq p} \sigma^3_{nn} \ket{u_n} \bra{u_n} \sigma^3 \Big) \ket{\partial_{k_a} u_p} &= - \sum_{n \neq p} \frac{\sigma^3_{nn}}{\bar{\omega}_{np}} v^a_{np} \ket{{n_n}} \notag\\
\Big( \mathcal{I} - \sigma^3_{pp} \ket{u_p} \bra{u_p} \sigma^3 \Big) \ket{\partial_{k_a} u_p} &= - \sum_{n \neq p} \frac{\sigma^3_{nn}}{\bar{\omega}_{np}} v^a_{np} \ket{u_n} \notag\\
\Big( \partial_{k_a} + i \mathcal{R}^a_{pp}  \Big) \ket{u_p} &= - \sum_{n \neq p} \frac{\sigma^3_{nn}}{\bar{\omega}_{np}} v^a_{np} \ket{u_n} ~.
\end{align}
By applying a gauge choice $ \mathcal{R}^a_{pp} = 0 $, we get
\begin{align}
    \ket{\partial_{k_a} u_p} &= \sum_{n \neq p} \frac{\sigma^3_{nn}}{\bar{\omega}_{pn}} v^a_{np} \ket{u_n}~, \\
    \bra{\partial_{k_a} u_p} &= \sum_{n \neq p} \frac{\sigma^3_{nn}}{\bar{\omega}_{pn}} v^a_{pn} \ket{u_n}~.
\end{align}
Using these identities in Eq. \eqref{L_a_np}, we get
\begin{align}
        L^a_{np} &= \frac{i}{4 \hbar} \epsilon_{abc} \sum_{\substack{m \neq n \\ q \neq p}} (\bar{\varepsilon}_n + \bar{\varepsilon}_p - 2 \bar{\varepsilon}_q) \frac{\sigma
        ^3_{mm} \sigma^3_{qq} \sigma^3_{mq}}{\bar{\omega}_{nm} \bar{\omega}_{pq} } v^b_{nm} v^c_{qp}  \notag\\
        &= \frac{i}{4} \epsilon_{abc} \sum_{m \neq n,p} \sigma^3_{mm} \Big( \frac{1}{\bar{\omega}_{nm}} + \frac{1}{\bar{\omega}_{pm}} \Big) v^b_{nm} v^c_{mp} .\label{orbital_moment_elements}
\end{align}
%
%

The electric dipole moment operator for the magnons can be written as \cite{Neumann_2024}
\begin{align}
    \hat{p}^a &= \frac{g \mu_B}{2c^2} ( \hat{\bm v} \times \hat{\bm m} - \hat{\bm m} \times \hat{\bm v} ),
    \\ \nn &= \frac{g \mu_B}{2 c^2} \epsilon_{abc} ( \hat{v}^b \sigma^3 \hat{s}^c + \hat{s}^c \sigma^3 \hat{v}^b )~.
\end{align}
In our work, we consider the collinear antiferromagnetic model, for which the $z-$component of spin is conserved. Using this property, we can simplify the matrix element of the electric dipole moment operator as follow:
\begin{align}
    p^a_{np} &= \frac{g \mu_B}{2c^2} \epsilon_{abz} \sum_m \sigma^3_{mm} [ v^b_{nq} s^z_{pp} \delta_{qp} + v^b_{qp} s^z_{qq} \delta_{nq} ] \notag\\
    &= \frac{g \mu_B}{2c^2} \epsilon_{abz} v^b_{np} [ \sigma^3_{pp} s^z_{pp} + \sigma^3_{nn} s^z_{nn} ]~.
\end{align}
The intraband matrix elements are given by $ p^a_{nn} = (g \mu_B/c^2) \epsilon_{abz} v^b_{np} \sigma^3_{nn} s^z_{nn}  $.

\bibliography{references}

\end{document}